\documentclass[%
 aip,
 sd,%
 amsmath,amssymb,
reprint,%
]{revtex4-1}

\usepackage{graphicx}
\usepackage{dcolumn}
\usepackage{bm}
\usepackage[english]{babel}
\usepackage[T1]{fontenc}
\usepackage[utf8]{inputenc}
\begin{document}

\preprint{AIP/123-QED}

\title[]{A Digital Matched Filter for Reverse Time Chaos}

\author{J. Phillip Bailey}
\email{bailejp@auburn.edu.}
\author{Aubrey N. Beal}
\author{Robert N. Dean}
\author{Michael C. Hamilton}%
\affiliation{ 
	Electrical and Computer Engineering Department, Auburn University, Auburn University, AL 36489
}%
\date{\today}

\begin{abstract}
The use of reverse time chaos allows the realization of hardware chaotic systems that can operate at speeds equivalent to existing state of the art while requiring significantly less complex circuitry. Matched filter decoding is possible for the reverse time system since it exhibits a closed form solution formed partially by a linear basis pulse. Coefficients have been calculated and are used to realize the matched filter digitally as a finite impulse response filter. Numerical simulations confirm that this correctly implements a matched filter that can be used for detection of the chaotic signal. In addition, the direct form of the filter has been implemented in hardware description language and demonstrates performance in agreement with numerical results.
\end{abstract}

\pacs{Valid PACS appear here}
\keywords{Suggested keywords}
\maketitle

\begin{quotation}
Recent developments in the field of chaos have shown that certain chaotic systems exhibit closed form analytic solutions. Because of the unique form of these solutions, it has been shown that a matched filter can be designed to detect the presence of chaotic waveforms generated by these exactly solvable systems. Hardware has been developed realizing both the chaotic system (in the form of a chaotic oscillator) and the matched filter, but both of these have been limited to relatively low frequencies. Reverse time chaos has been presented as a means to increase the achievable operating frequency by redefining the chaotic system equations in a form that allows a hybrid digital and analog circuit realization. This work expands on reverse time chaos by deriving a numerical model of a matched filter for the reverse time system. In addition, a digital FIR implementation of this filter is demonstrated to function correctly in simulation.
\end{quotation}

\section{\label{sec:level1}Introduction}
Although many types of chaos continue to evade analytic solutions, the assumption that no solvable chaotic dynamics exist has been broken by the discovery of hybrid (consisting of interdependent continuous time and discrete time states) chaotic systems that exhibit exact solutions \cite{umeno1997method,corron_matched_2010,corron_exact_2012}. A linear matched filter, which can be used to exactly recover the discrete-time state of the chaotic system when driven with only that system's continuous-time state, has also been developed using these solutions. Matched filter decoding significantly reduces the difficulty of integrating chaos into a wide range of electronic systems, including both wireless communication and radar.

One of the most promising areas of research in chaotic systems has long been in their utility for communication \cite{grebogi_communicating_1993,hayes_chaos_2005}. Previous work has shown that such systems can be constructed and tuned such that they possess advantageous dynamics for various communication schemes \cite{heidari1994chaotic,hirata_constructing_2005}. In addition to possessing these targeted dynamics, chaotic systems can also be subjected to control schemes that can maintain desired trajectories \cite{hayes_experimental_1994}. By combining these characteristics with matched filter decoding \cite{corron_matched_2010,cohen_pseudo-matched_2012}, many of the necessary components for a modern high-performance communication system may be realized with chaos-based systems. 

Various elements of such systems have been analyzed to determine how they might perform in a number of applications. In the presence of additive Gaussian white noise, chaotic systems can exhibit error rates comparable to communication systems currently in use \cite{chen_stochastic-calculus-based_2000,blakely_communication_2013}. Many issues associated with multipath propagation may also be mitigated when the propagated waveform is chaotic \cite{kolumban_theoretical_2000}. 

Reverse time chaos provides a potential solution for realizing chaos in hardware with both solvable and controllable properties without sacrificing the ability to scale in speed. First proposed by Corron {\it et~al.} in \cite{corron_chaos_2006}, reverse time chaos describes behavior that differs from traditional chaos by using the current state of the system to represent all of its past states instead of all of its future states. Despite this difference, reverse time chaos retains a positive Lyapunov exponent and a corresponding sensitivity to initial conditions that defines traditional chaotic systems. The many similarities between traditional (or forward time) and reverse time chaos also allow for formulation of an exact solution. Because of this, in a method analogous to that used for the forward time systems in \cite{corron_matched_2010,corron_exact_2012}, the development of a linear matched filter that can be used to detect the original chaotic signal is also possible.  

\pagebreak
In this paper, we build on recent work demonstrating a functional reverse time chaotic oscillator \cite{bailey_high-frequency_2014} by developing an expression for its corresponding matched filter. Realization of the matched filter completes the transmitter and receiver system presented in previous work using forward time chaos discussed above. With this complete system, many of the advantages cited for the forward time case can also be realized with reverse time chaos.

\section{\label{sec:chaos}Exactly Solvable Chaos}
\subsection{Forward Time}
The chaotic system used by Corron {\it et~al.} can be described by a set of two equations written in terms of two states. These consist of a continuous time state $u(t) \in \mathbb{R}$ that provides exponential growth so that the system does not die out and a discrete time state $s(t) \in \{-1, 1\}$ or $s(t) \in \{0, 1\}$ that maintains boundedness so that the system does not grow unbounded. In order to differentiate its behavior from the reverse time system discussed in the next section, the dynamics described by this behavior are those that have been previously referred to as forward time chaos. $u(t)$ is defined by the continuous-time differential equation:
\begin{equation} \label{ctdeq}
\ddot{u} - 2\beta\dot{u} + (\omega^2+\beta^2)*(u-s) = 0
\end{equation}
where $\omega = 2\pi$ and $0<\beta\leq\ln{2}$. Various system dynamics can be realized by selecting specific definitions of $s(t)$; for Lorenz-like dynamics, $s(t)$ is defined as:
\begin{equation} \label{sbsw}
\dot{u} = 0 \rightarrow s(t) = sgn(u(t))
\end{equation}
where
\begin{equation} \label{sgn}
sgn(u) = \begin{cases}
-1 & u<0\\
1 & u\geq0
\end{cases}
\end{equation}

\begin{figure}[ht!]
	\begin{center}
\includegraphics[width=0.4\textwidth,height=0.2\textwidth]{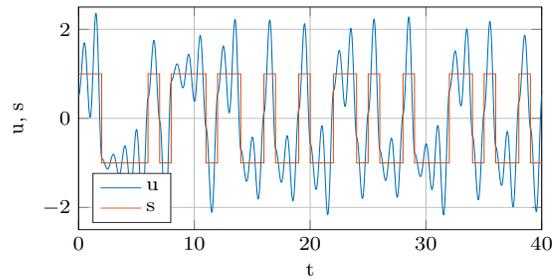}
	\end{center}
	\caption{Numerical simulation of exactly solvable system in shift band (time evolution)}\label{ft_shift_ty}
\end{figure}

When operating with this $s(t)$, the system is said to be in the shift band. Figure \ref{ft_shift_ty} demonstrates this behavior graphically.
\pagebreak

\begin{figure}[ht!]
	\begin{center}
		\includegraphics[width=0.4\textwidth,height=0.2\textwidth]{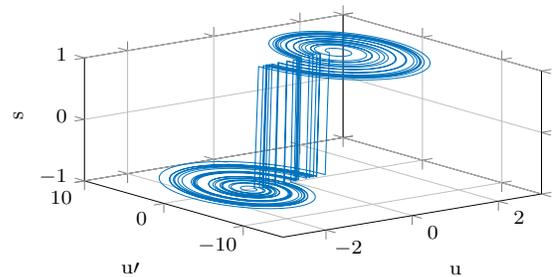}
	\end{center}
	\caption{Numerical simulation of exactly solvable system in shift band (phase portrait).}\label{ft_shift_3d}
\end{figure}

As with many other chaotic systems, the phase portrait of this system provides more insight into long term dynamics. The phase portrait for the shift band is presented in Figure \ref{ft_shift_3d}.

Although the system dynamics are captured by $u$ and $\dot{u}$, $s$ is also included to create a three-dimensional plot; this allows for an alternate visualization of the system oscillations $u$ being centered around the discrete levels of $s$. Despite the complexity this system appears to possess when viewed graphically, it can be shown to have an analytic solution given as:
\begin{equation}\label{sol_ft}
u(t)=\sum_{m=-\infty}^{\infty}s_m P(t-m)
\end{equation}
\begin{equation} \label{basis_ft}
P(t) = \begin{cases}
(1-e^{-\beta})e^{\beta t}(\cos(\omega t)-\frac{\beta}{\omega}\sin(\omega t)) & t<0\\
1-e^{\beta(t-1)}(\cos(\omega t)-\frac{\beta}{\omega}\sin(\omega t)) & t=0\\
0 & t>0
\end{cases}
\end{equation}
where $P(t)$ acts as a basis function and $s_m\in{-1,1}$ is the symbol sequence defined for all $t$\cite{corron_matched_2010}. Although the simple existence of this solution is remarkable, its most significant characteristic is in its description in Equation \ref{sol_ft} in terms of convolution of a basis function with a random bipolar pulse. This description notably allows for derivation of the matched filter that can be used to partially regenerate $s(t)$ when presented with only $u(t)$. The matched filter can be described by:
\begin{equation}\label{mf_ft_s1}
\ddot{\xi}+2\beta\dot{\xi}+(\omega^2+\beta^2)\xi=(\omega^2+\beta^2)\eta(t)
\end{equation}
where
\begin{equation}\label{mf_ft_s2}
\dot{\eta}=v(t+1)-v(t)
\end{equation}
In Equations \ref{mf_ft_s1} and \ref{mf_ft_s2}, $\xi$ represents the output of the matched filter (which is also the recovered $s(t)$) while $v(t)$ represents the input, which is assumed to be $u(t)$ \cite{corron_matched_2011}. Previous work has demonstrated that hardware implementations of this filter can be used to successfully recover the original $s(t)$ \cite{corron2013acoustic}.
\subsection{Reverse Time}
Although both components of the forward time exactly solvable system have been extended to circuit simulations operating near 2 MHz \cite{beal_design_2012}, published hardware implementations operate in the audio frequency range \cite{corron_matched_2010,bailey2012simulation}. Each circuit consists of many analog components that scale poorly with frequency and do not integrate well with modern high density integrated circuit (IC) processes. In particular, the use of a negative impedance converter in many of the published circuits provides a significant challenge for high frequency designs of the oscillator and the requirement for a long delay time adds significant complexity to the matched filter.

Reverse time chaos provides a method for generating functionally equivalent waveforms with significantly simpler circuit implementations than have been published for the forward time case \cite{corron_matched_2010,beal_design_2012}. In contrast to these circuits, which use many active components that may be difficult to scale in frequency, the reverse time chaotic oscillator requires only three passive components (forming a series resistor, inductor, and capacitor - RLC -  tank) in its simplest form; the remainder of the circuit can be implemented digitally \cite{bailey_high-frequency_2014}. Complementing this, the reverse time matched filter presented in this work can be implemented entirely in digital circuitry. 

A mathematical description of reverse time chaos can be defined by the second order system:
\begin{equation}\label{diffeq_main}
\ddot{u}+2\beta\dot{u}+(\omega^2+\beta^2)u=(\omega^2+\beta^2)s(t)
\end{equation} 
as first described in \cite{corron_chaos_2006}. Both $\beta$ and $\omega$ are control parameters for this system; however, to maintain consistency with typical engineering practice, $\omega$ will be defined as the radian frequency of operation. $s(t)$, the forcing function, can be described as
\begin{equation}\label{s_main}
s(t) = As_n
\end{equation}
where $s_n\in \{-1, 1\}$ is a random sequence with amplitude $A$, which can take any real value. 

\begin{figure}[ht!]
	\begin{center}		
\includegraphics[width=0.4\textwidth,height=0.2\textwidth]{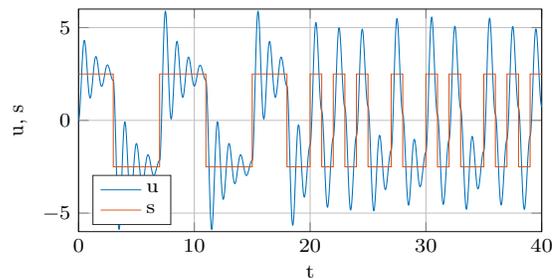}	
	\end{center}
	\caption{Numerical simulation of reverse time system in shift band (time evolution).}\label{rt_shift_ty}
\end{figure}

Upon inspection, it is apparent that Equation \ref{diffeq_main} describes a harmonic oscillator with positive damping producing oscillations centered around the value of $s(t)$. An illustration of this behavior generated with numerical simulation for $\beta=\mathrm{ln}(2)$ and $\omega=2\pi$ is shown in Figure \ref{rt_shift_ty}.

The positive damping term causes the value of $u(t)$ to decrease in magnitude and is the reason that the reverse time system requires $s(t)$ to act as a forcing function. Without this forcing, $u(t)$ would simply die out after ringing down to 0. 

As described in the introduction, a reverse time chaotic system differs from a forward time chaotic system by using its initial condition to describe the set containing all of its previous states rather than its future states. To better visualize this, Figure \ref{rt_shift_nty} below shows Figure \ref{rt_shift_ty} with its x axis reversed.
\begin{figure}[ht!]
	\begin{center}
\includegraphics[width=0.4\textwidth,height=0.2\textwidth]{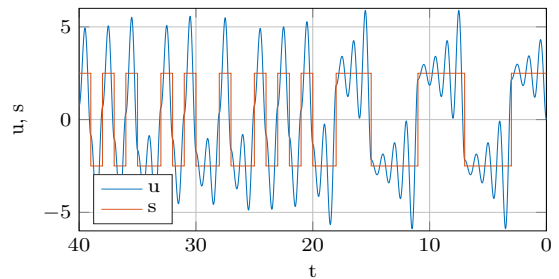}	
	\end{center}
	\caption{Numerical simulation of reverse time system in shift band (time reversal).}\label{rt_shift_nty}
\end{figure}

\begin{figure}[ht!]
	\begin{center}
\includegraphics[width=0.4\textwidth,height=0.2\textwidth]{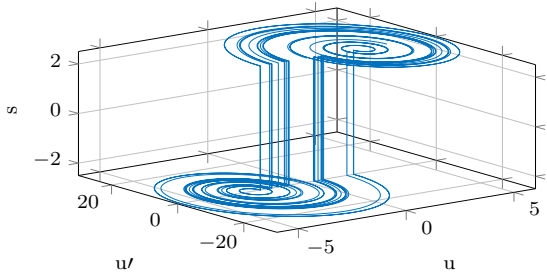}	
	\end{center}
	\caption{Numerical simulation of reverse time system in shift band (phase portrait).}\label{rt_shift_3d}
\end{figure}

When viewed in this manner, the dynamics of $u(t)$ appear identical to those of \ref{ctdeq} operating in the shift band (Figure \ref{ft_shift_ty}). Further confirmation of the similarity is shown by examining the phase space projection, shown in Figure \ref{rt_shift_3d}.\pagebreak

The reverse time system's phase portrait matches almost exactly with the phase portrait from the forward time system apart from the levels of $s$ around which the trajectories orbit. In the previous section, these were -1 and 1, while in Figure \ref{rt_shift_3d} they are -2.5 and 2.5, which corresponds to the value of $A$ used in the reverse time numerical analysis routine. As in the forward time system, a solution can be described in terms of a basis pulse; the basis pulse for the reverse time system is given as:

\begin{equation}\label{rt_basis}
u_g(t)=\begin{cases}
0 & t < 0\\
\frac{1-e^{-\beta t}}{\omega^2+\beta^2}\left[(\cos(\omega t)+(\frac{\beta}{\omega})\sin(\omega t))\right] & 0\leq t < 1\\
\frac{e^{-\beta t}}{\omega^2+\beta^2}\left[(\cos(\omega t)+(\frac{\beta}{\omega})\sin(\omega t))\right] & t\geq1
\end{cases}
\end{equation}

\begin{figure}[ht!]
	\begin{center}
\includegraphics[width=0.4\textwidth,height=0.2\textwidth]{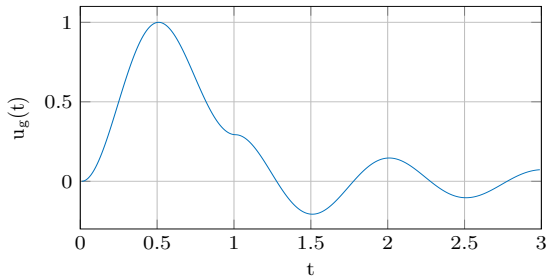}	
	\end{center}
	\caption{Reverse time basis pulse.}\label{mf_basis}
\end{figure}

A plot of Equation \ref{rt_basis} evaluated for $0 \leq t < 3$ is shown in Figure \ref{mf_basis}.

The difference between the two cases in Equation \ref{rt_basis} is clearly visible; for $0 \leq t < 1$, the basis pulse is offset by 1 while for $t \geq 1$, this offset is removed. As expected, this response is identical but reversed in time to the forward time basis pulse in Equation \ref{basis_ft}. Because the system's solution can be described in terms of this basis pulse, it can be extended to all $t$ \cite{corron_chaos_2006}:
\begin{equation}\label{rt_final_sol}
u(t)=\sum_{n=-\infty}^{t}s_n u_g(t-n)
\end{equation}

The existence of this solution also allows for the development of a matched filter with the same process used for the forward time system. 

\section{Matched Filter}
A common problem in many signal processing applications is to determine if a known signal is present in a received signal that may contain significant noise. Although there are multiple approaches to solve this problem, the use of a filter whose response is matched to the known signal - a matched filter - is frequently the solution of choice. This matched filter can be described as the system with the maximal output SNR when presented with a known signal corrupted by additive white Gaussian noise (AWGN). A brief mathematical description follows \cite{fsig,haykin_communication_2001}.

\subsection{Derivation}
Given a known transmitted signal $x(t)$, the received signal $v(t)$ can be written as:
\begin{equation}
v(t)=Ax(t-t_p)+n(t)
\end{equation}
where $A$ is amplitude scaling due to attenuation or gain of the channel, $t_p$ is the propagation delay through the channel, and $n(t)$ is AWGN added in the channel with a power spectral density (PSD) $S_n(\omega)$. The SNR of this system can be defined as:
\begin{equation}
\mathrm{SNR}=\frac{\max{(|Ax(t-t_0)|^2)}}{\overline{|n(t)^2|}}=\frac{|Ax(0)|^2}{\overline{|n(t)|^2}}
\end{equation}

The matched filter is the linear filter ($h(t)$ or $H(\omega)$) that maximizes its output signal-to-noise ratio (SNR) at some instant relative to a starting point $t_0$ and can be can be written as \cite{haykin_communication_2001}:
\begin{equation}\label{mf_final}
h(t)=kx(t_x-t)
\end{equation}

\begin{figure}[ht!]
	\begin{center}
\includegraphics[width=0.4\textwidth,height=0.2\textwidth]{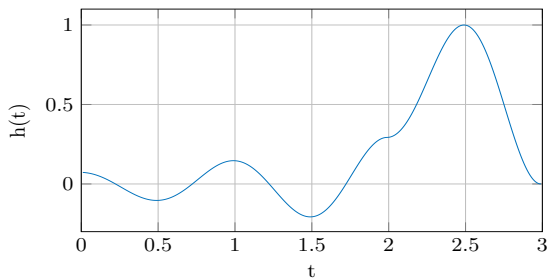}	
	\end{center}
	\caption{Reverse time matched filter response.}\label{mf_basis_f}
\end{figure}

Equation \ref{mf_final} describes a filter whose response is identical to the known transmitted signal $x(t)$ but reversed in time and shifted by some amount $t_x$ ($k$ may be used to adjust the amplitude, but typically is left equal to 1). For the reverse time system, the received response is taken to be the basis pulse given by Equation \ref{rt_basis}. The resulting required matched filter response is shown in Figure \ref{mf_basis_f} for $\beta=\ln(2)$.  \pagebreak

\subsection{Verification}
\begin{figure}[ht!]
	\begin{center}
\includegraphics[width=0.4\textwidth,height=0.2\textwidth]{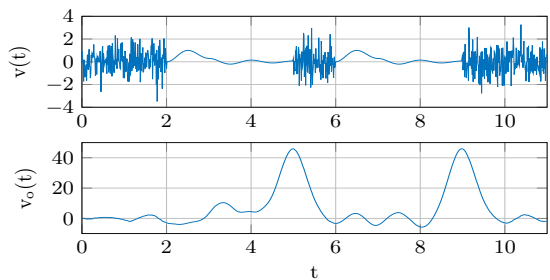}	
	\end{center}
	\caption{Matched filter response to basis pulse. (top) Input to filter and (bottom) output from filter.}\label{mf_test_b}
\end{figure}

Using a numerical model for the matched filter, its performance in a number of scenarios of interest was evaluated. The first test performed was to determine if the filter could properly detect the basis pulse when inserted into a random signal. Figure \ref{mf_test_b} demonstrates this process.

\begin{figure}[ht!]
	\begin{center}
\includegraphics[width=0.4\textwidth,height=0.2\textwidth]{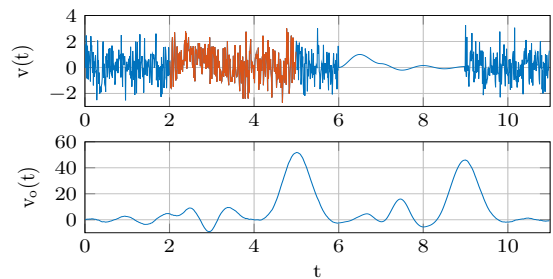}	
	\end{center}
	\caption{Matched filter response to basis pulse in presence of AWGN (SNR=0 dB). (top) Input to filter with basis pulse overlaid in red and (bottom) output from filter.}\label{mf_test_b_n0}
\end{figure}

The input ($v(t)$) is shown in the top while the output from the numerical filter ($v_o(t)$) is shown in the bottom plot. These results indicate that it correctly responds to the presence of a basis pulse. To verify the expected resilience in the presence of AWGN, three additional tests were performed with increasing noise levels added for each successive run. MATLAB's {\tt awgn} function was used to add noise to the existing model. Figure \ref{mf_test_b_n0} shows these results with the SNR set to 0 dB. In this figure, the segment containing the basis pulse with added noise is highlighted in red.\pagebreak

Despite the significant noise added to the first instance of the basis pulse, the matched filter was able to correctly detect its presence with approximately the same response magnitude as generated by the basis pulse alone. The response to the random segments at the beginning and end of the input signal remained low throughout; this result combined with the peak behavior previously discussed indicates a strong selectivity for the basis pulse. 

\subsection{Chaotic Input/Matched Filter Output}
\begin{figure}[ht!]
	\begin{center}
\includegraphics[width=0.4\textwidth,height=0.2\textwidth]{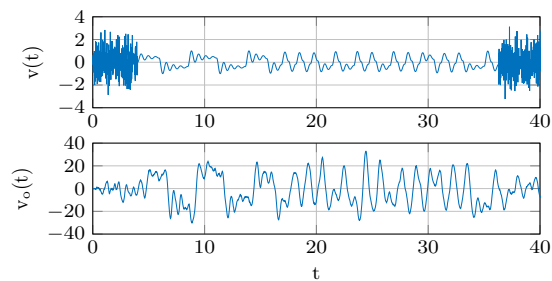}	
	\end{center}
	\caption{Matched filter response to reverse time chaotic waveform. (top) Input to filter and (bottom) output from filter.}\label{mf_chaos_in}
\end{figure}

While the ability to detect a single basis pulse provides verification that the matched filter model is operating correctly, its true utility comes from the ability to detect a reverse time chaotic signal. This signal consists of many instances of the basis pulse summed together, therefore the matched filter output shows a large number of peaks as the input is shifted through its response. The magnitude of these peaks is diminished as compared to the response when presented with a single instance of the basis pulse, but will be shown in the next section to be high enough to allow for reliable detection of the chaotic waveform. Reusing the numerical model from the previous section, a chaotic signal was inserted into a random waveform to verify the filter's functionality. The output of this test is shown in Figure \ref{mf_chaos_in}. 

\begin{figure}[ht!]
	\begin{center}
\includegraphics[width=0.4\textwidth,height=0.2\textwidth]{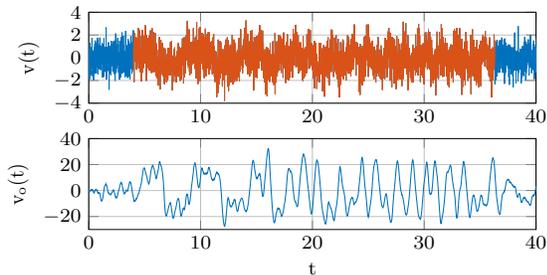}	
	\end{center}
	\caption{Matched filter response to reverse time chaotic waveform in presence of AWGN (SNR=0 dB). (top) Input to filter with chaotic signal overlaid in red and (bottom) output from filter.}\label{mf_chaos_in_0}
\end{figure}

For completeness, the effect of AWGN on the chaotic waveform has also been explored with the numerical matched filter. The same noise level (SNR = 0 dB) used peviously was selected for this analysis. These results are presented in Figure \ref{mf_chaos_in_0}.

As with the basis pulse, even high magnitude AWGN does not significantly diminish the response of the matched filter to the presence of a chaotic waveform.

\subsection{$s_n$ Reconstruction}
\begin{figure}[ht!]
	\begin{center}
\includegraphics[width=0.4\textwidth,height=0.2\textwidth]{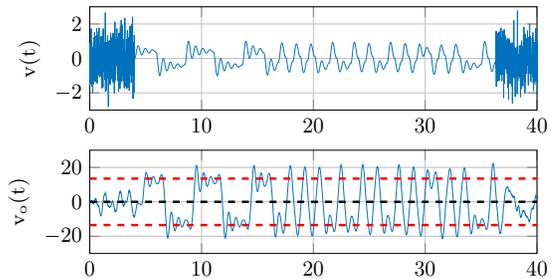}	
	\end{center}
	\caption{Threshold levels used for reconstruction of $s_n$. (top) Input to filter and (bottom) output from filter with threshold levels indicated. }\label{mf_thresh}
\end{figure}

Limited post-processing can be applied to the matched filter response to reconstruct the $s_n$ sequence used to generate the chaotic signal applied to its input. This post-processing method requires the establishment of three threshold levels: a high threshold, a zero crossing or midpoint threshold, and a low threshold. These thresholds were determined by visual inspection in this work; although non-optimal, this method still allows for proper reconstruction. Figure \ref{mf_thresh} demonstrates the placement of these levels where the high and low thresholds are denoted by the upper and lower red lines and the midpoint threshold is denoted by the black line.\pagebreak 

\begin{figure}[ht!]
	\begin{center}
\includegraphics[width=0.4\textwidth,height=0.2\textwidth]{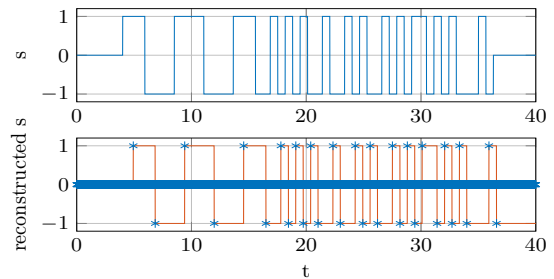}	
	\end{center}
	\caption{Reconstructed $s_n$. (top) Original $s$ signal from oscillator and (bottom) reconstructed $s$ from filter output with each decision point indicated by a blue asterisk.}\label{mf_rec}
\end{figure}

New values of $s_n$ always correspond to the matched filter output going above (or below) the high and low threshold values. Simply detecting these events, however, does not properly reconstruct $s_n$ - it only indicates the presence of the basis pulse (it is important to reiterate that the matched filter is matched to the basis pulse, not the overall chaotic waveform or $s_n$). The solution in Equation \ref{rt_final_sol} allows for $s_n$ to hold its value for many cycles, which results in many oscillations around the threshold. To account for this, the midpoint threshold is used to detect when a switch in the $s_n$ value has occurred and then update the recovered $s_n$. This process is shown graphically in Figure \ref{mf_rec}.

When compared to the original $s$ state of the reverse time chaotic oscillator (top), the reconstructed $s$ (bottom) appears identical apart from the time shift introduced by the matched filter operation. The blue asterisks marked on the reconstructed $s$ plot indicate the decision made by the post processing algorithm at each time step. For the vast majority of the steps, this value is 0, corresponding to no change in $s$ (i.e., many asterisks are repeated at reconstructed $s=0$). When the matched filter output crosses through the midpoint threshold and then exceeds either the high or low threshold, however, the decision changes to indicate a new value. These decision points are then used to generate the overall reconstructed $s$ as shown.

\section{FIR Implementation}
Numerical modeling in MATLAB provided a straightforward path for realization of the matched filter in hardware as a digital FIR filter. The selection of a digital filtering method provides two distinct advantages: it allows for a highly precise definition of the filter response while also side stepping the challenges raised by attempting to implement the inherently unstable reverse time matched filter response in analog hardware. Values used for the filter's magnitude response can be used directly to define the coefficients necessary for the FIR algorithm. Any N\textsuperscript{th} order constant coefficient digital FIR function can be described as a convolution sum as shown in Equation \ref{xn} \cite{fir_fpga}.

\begin{equation} \label{xn}
Y[n] = X[n] * H[n] = \sum_{k = 0}^{N - 1} H[k] * X[n - k]
\end{equation}

The exact function performed is determined by the values of $H$, which are the FIR coefficients. This form can be implemented in hardware directly (and is often referred to as the direct-form FIR), but is generally less desirable due to its use of multipliers. One solution for addressing this shortcoming is using the sum of power of two (SOPOT) method to decompose multiplication into addition and multiplication by powers of two \cite{sopot,matlab}. Equation \ref{sop} describes the process of determining the SOPOT representation of an arbitrary value $s(n)$. 
\begin{equation} \label{sop}
s(n) = \sum_{i = 0}^{J - 1} a_{i,n} * 2^{b_{i,n}}
\end{equation}
where $a_{i,n} \in \{-1, 0, 1\}$, $b_{i,n} \in \{0, 1,..., u\}$, and $J$ is the number of terms necessary to represent $s(n)$ \cite{sopot}. The maximum value that can be represented by SOPOT is related to $u$, so its value determines the allowable range of coefficients. Because SOPOT contains only multiplication by powers of two, the multiplication functions can be replace by logical left shifts that shift by the $b_{i,n}$ values. An algorithm based on the work in \cite{dhobi_fpga_2014} was developed in MATLAB to automate the calculation of these coefficients for any real number.

\subsection{Architecture}
\begin{figure}[ht!]
	\begin{center}
		\includegraphics[width=.4\textwidth]{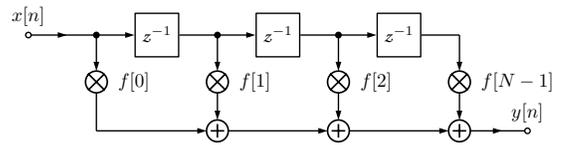}
	\end{center}
	\caption{FIR filter block diagram.}
	\label{fir}
\end{figure}

A general block diagram of the FIR architecture is shown in Figure \ref{fir}\cite{damian_low_2011}.\pagebreak

From this figure, it is apparent that the algorithm requires relatively few components for its hardware realization. The unit delay blocks (indicated as $z^{-1}$) map directly to D flip-flops where the clock period sets the delay time. The weights (denoted by $f[0]$ to $f[N-1]$) require a varying number of adders and left shift circuits as indicated by the SOPOT decomposition. A MATLAB routine was also written to generate and combine these blocks automatically, given a set of FIR weights.

\subsection{HDL Simulation}
\begin{figure}[ht!]
	\begin{center}
\includegraphics[width=0.4\textwidth,height=0.4\textwidth]{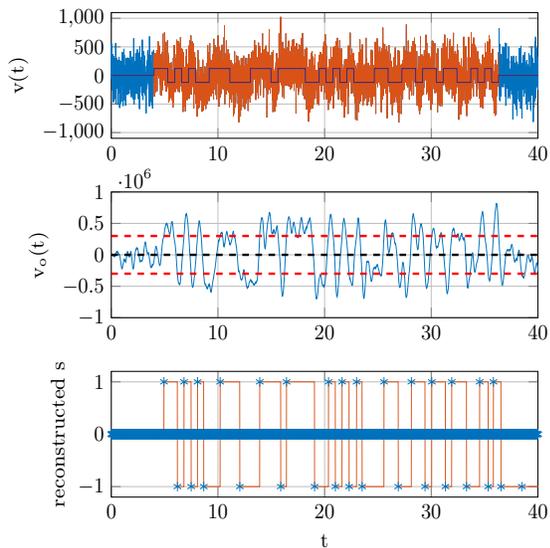}	
	\end{center}
	\caption{Simulation results of FIR matched filter with AWGN (SNR=0 dB). (top) Input to filter with chaotic signal overlaid in red and original $s$ signal, (middle) output from filter with threshold levels indicated, and (bottom) reconstructed $s$ from filter output with each decision point indicated by a blue asterisk. }\label{real_noise}
\end{figure}

Making use of the Verilog code generated for the FIR filter, a HDL simulation was performed using ModelSim to determine expected performance of a physical implementation. 32 bit values were used for internal calculations to allow for high precision while 10 bit values were used for the input to correspond to typical ADC resolutions. An operating clock frequency was set to 180 MHz due to the use of the 1.8 MHz chaotic oscillator with 100 FIR coefficients. A simulation, shown in Figure \ref{real_noise}, confirms the code's performance in the presence of AWGN.\pagebreak

As expected, the Verilog model is able to correctly detect the chaotic signal despite the significant magnitude of added noise. The large scales used for the y axis on both the filter input and output result from converting the binary values generated from the simulation into decimal directly. 

\subsection{System Performance}
With the reconstructed $s$ sequence, performance of the oscillator and matched filter combination can be explored. A common method of evaluating communication systems is to examine their bit error rate (BER), which can be described by:
\begin{equation}\label{BER_an}
\mathrm{BER}=\frac{\mathrm{number\ of\ incorrect\ bits}}{\mathrm{number\ of\ total\ bits}}
\end{equation}

\begin{figure}[ht!]
	\begin{center}
		\includegraphics[width=0.4\textwidth,height=0.2\textwidth]{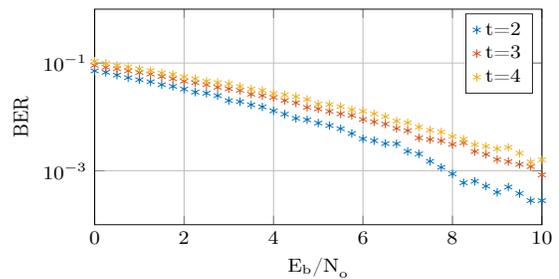}	
	\end{center}
	\caption{Simulated BER for basis pulses of varying length.}\label{BER_bplen}
\end{figure}

Evaluating the BER for varying lengths (maximum values of $t$) of the basis pulse in Equation \ref{rt_basis} allows for selection of the optimal length. Figure \ref{BER_bplen} demonstrates the effects of varying $t$ from 3 to 5.\pagebreak

\begin{figure}[ht!]
	\begin{center}
		\includegraphics[width=0.4\textwidth,height=0.25\textwidth]{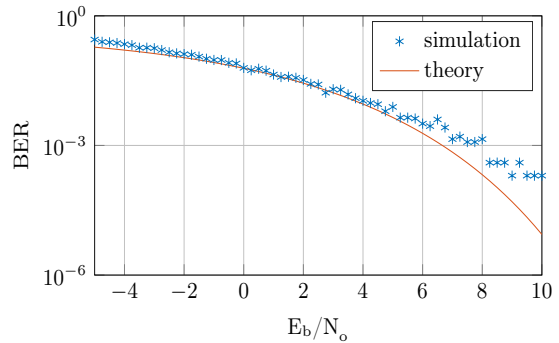}	
	\end{center}
	\caption{Comparison of theoretical and simulated BER.}\label{BER}
\end{figure}

From the above results, it is immediately apparent that the best performance is realized with $t=3$. This result was selected for a comparison against the previously published theoretical performance \cite{corron_matched_2010}. Results of the comparison are shown in Figure \ref{BER}. 

While the theoretical BER takes in to account the error induced by the determinism inherent in a chaotic waveform, it does not address addition errors introduced by the digitization and truncation of the matched filter that was performed in this work. As expected, with these included, the performance of the overall system is slightly reduced as compared to theoretical expectations. 

\section{Conclusions}
Matched filter decoding has been shown to provide a viable means of detection of reverse time chaotic waveforms. Building on previous work, a numerical model was constructed and analysis predicted that these waveforms could be reliably detected even in the presence of high magnitude AWGN. Both this ability and the ability to reconstruct the original $s$ sequence using a simple post-processing scheme have been demonstrated. 

Conversion of the numerical model to a FIR filter has been performed and implemented in direct form. To more readily translate to a synthesizable design, this form was decomposed using the SOPOT method and translated into HDL code. Simulation results of this code indicate that a hardware realization is straightforward and practical on both a low cost FPGA and a high performance ASIC. 

Finally, these results confirm that reverse time chaos can be used in the same transmitter/receiver model as forward time chaos without requiring the complex and difficult to scale hardware associated with existing forward time chaotic systems. Applications in many areas including communications and radar can be satisfied with the simple hardware required for reverse time chaos. 

\begin{acknowledgments}
The authors wish to acknowledge Dr. Lloyd Riggs for providing insight on the mathematics and operation of matched filters.
\end{acknowledgments}

\bibliography{fir_rt}

\end{document}